\documentclass[12pt,preprint,aps]{revtex4}

\usepackage{bm}% bold math
\usepackage{graphicx}
\usepackage{epsfig}
\begin{document}

\preprint{IZTECH-P08-2005}

\title{Reconsidering extra time-like dimensions}

\author{Recai Erdem and Cem S. \"{U}n }
\email{recaierdem@iyte.edu.tr,  cemun@iyte.edu.tr}
\affiliation{Department of Physics,
{\.{I}}zmir Institute of Technology \\ 
G{\"{u}}lbah{\c{c}}e K{\"{o}}y{\"{u}}, Urla, {\.{I}}zmir 35430, 
Turkey} 

\date{\today}

\begin{abstract}
In this study we reconsider the phenomenological problems related to 
tachyonic modes in the context of extra time-like dimensions. First we 
reconsider a lower bound on the size of 
extra time-like dimensions and improve the conclusion in the literature. 
Next 
we discuss the issues of spontaneous decay of stable fermions through 
tachyonic decays and disappearance of fermions due to 
tachyonic contributions to their self-energies. We find 
that the tachyonic modes due to extra time-like dimensions are less 
problematic than the tachyonic modes in the usual 4-dimensional setting 
because the most troublesome Feynman diagrams are forbidden once 
the conservation of momentum in the extra time-like dimensions is imposed. 
\end{abstract}

\maketitle

Extra spatial coordinates are considered thoroughly in recent years. A 
glance at ArXiv shows that there are hundreds of papers on extra 
dimensions in the last five years and almost all of them being 
wholly or mainly on spatial extra dimensions. From the theoretical 
point of view the scarcity of studies 
involving extra time-like dimensions \cite{Old,Gabadadze,Chaichian,Li,
Iglesias, Berezhiani,Erdem} 
is mainly due to the existence of 
tachyonic modes in such models, which are problematic because of the 
violation of causality 
and unitarity and lack of an adequate field theoretic description of 
tachyonic fields \cite{Gabadadze} while from the phenomenological point 
of view the most serious problems are the extremely small empirical lower 
bound in literature on the size(s) of extra time-like dimensions 
\cite{Yndurain}, 
the spontaneous decay of 
stable particles induced by negative energy tachyons 
\cite{Feinberg,Gabadadze}, the imaginary self energy for charged fermions 
induced by tachyonic photon modes, which in turn, seems 
to cause disappearance of the fermion into nothing in a very short time 
\cite{Gabadadze}. In 
this study we will focus 
on the phenomenological difficulties and try to seek if one may moderate 
the phenomenological problems mentioned above with the hope that 
a thorough consistent formulation of the field theory of tachyons and 
their 
interactions with the usual particles may be formulated in future
(if tachyons exist at all). The first phenomenological problem that will 
be considered here is the extremely small lower bound 
derived from the lower bound on the lifetime of proton 
\cite{Yndurain}.
In the light of this extremely small lower bound on the size of extra 
time-like dimension(s), in the order of a tenth of the Planck scale, 
either one should dare to employ such ( unnaturally) small dimension(s)or 
should use brane models where our physical world is a brane with an 
infinitesimal width in the extra time-like direction \cite{Gabadadze} or a 
scheme where tachyonic modes are not allowed to be produced 
\cite{Iglesias,Berezhiani}. 
A possible relaxation of the bound on the size of extra time 
dimension(s) would 
give more freedom to the model 
constructions with extra time-like dimension(s). So we reconsider the 
lower 
bound obtained from the lower bound on the proton lifetime and the 
calculation of a tree level Feynman diagram. We find that 
the calculation 
leads to no bound on the size of extra time-like dimensions. 
In fact we just repeat the calculations in
\cite{Yndurain} except we notice the fact that there is a cutoff 
momentum in the Fourier transform. In other words the difference 
between our result and the original study results 
from the naive 
application of the Fourier transform in \cite{Yndurain}
to get the non-relativistic potential 
corresponding to the scattering of protons inside a nucleus by tachyonic 
photon modes. In the original study 
the effect of tachyonic modes on fermion self-energies are neglected and
no cutoff was taken, the integration is from minus infinity to plus 
infinity in momenta while one should take a cutoff corresponding to the maximum 
momentum available to the protons inside the nucleus.
One obtains the same result as the one obtained in \cite{Yndurain} when 
one lets the cutoff momentum go infinity and neglects the self-energy 
contributions. Next we consider the problems of 
the spontaneous decay of the particles through release of negative energy 
tachyons and the imaginary mass induced through self energy diagrams of 
fermions. We argue that these problems may be evaded by imposing 
conservation of momentum in the extra time direction provided that the 
standard model particles are identified as the zero modes of the 
Kaluza-Klein tower ( that is the standard identification). 

First consider the following tree level diagram for the electromagnetic
scattering of two protons inside a nucleus \cite{Sakurai,Yndurain2}. 
\begin{figure}[h]
\vspace*{-0.65in}
\hspace*{-0.7in}
\begin{minipage}{4.50in}
\begin{center}
\epsfig{file=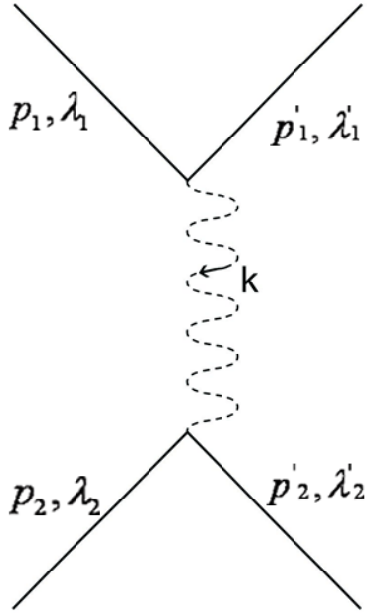,height=5.25in}
\end{center}
\end{minipage}
\vskip -0.65in
\caption{\label{fig1}
{\it
The Feynman diagram for the scattering of two protons with the
initial 4-momenta and the spins; $p_{1}$, $p_{2}$ and
$\lambda_{1}$, $\lambda_{2}$ and the final 4-momenta and the spins
$ p_{1}^{'}$, $p_{2}^{'}$ and $\lambda_{1}^{'}$,
$\lambda_{2}^{'} $.The wavy line denotes the tachyonic
Kaluza-Klein modes of photon}} 
\end{figure}
%\clearpage

The scattering cross section corresponding to this diagram may be obtained 
from
the scattering amplitude of elastic fermion-fermion scattering. The 
differential cross section for 
elastic fermion-fermion scattering is related to scattering
amplitude T by
\begin{eqnarray}
\frac{d\sigma}{d\Omega}=|T|^{2}=\frac{1}{2p_{10}2p_{20}2p^{'}_{10}2p^{'}_{20}}
|M|^{2} \label{e1}
\end{eqnarray}
where M is the matrix element given by
\begin{eqnarray}
M=\frac{e^{2}}{4\pi^{2}}\overline{u}_{(p^{'}_{1},\lambda^{'}_{1})}
\gamma_{\mu}u_{(p_{1},\lambda_{1})}
\,\frac{1}{k^{2}+m_{n}^{2}+i0}\,\overline{u}_{(p^{'}_{2},\lambda^{'}_{2})}
\gamma^{\mu}u_{p_{2},\lambda_{2})} \label{e2}
\end{eqnarray}
where
\begin{eqnarray}
m_{n}^{2}=\frac{n^{2}}{L^{2}}\label{e3}
\end{eqnarray}
and u's are 4-component Dirac spinors,$\gamma_{\mu}$ are gamma
matrices. One should also include the exchange scattering where
$p^{'}_{1}\leftrightarrow p^{'}_{2}$,
$\lambda^{'}_{1}\leftrightarrow\lambda^{'}_{2}$, but we are only
interested in the order of magnitude results and the crossed term
of (\ref{e2}) gives a similar contribution as (\ref{e2}) itself and does 
not alter the conclusion. So it is sufficient to consider (\ref{e2}). In 
the non-relativistic limit \cite{Yndurain2} the zero component of 
the proton 4-momenta $p_{01}$, $p_{02}$ and the photon 4-momentum transfer 
k are approximated by
\begin{eqnarray}
p_{0}\simeq
m+\frac{\vec{p}^{\,2}}{2m}-\frac{\vec{p}^{\,4}}{8m^{3}}
\nonumber
\end{eqnarray}
\begin{eqnarray}
k^{2}=(p_{1}-p_1^{\prime})^{2}=(p_{10}-p_{10}^\prime)^{2}
-(\vec{p}_{1}-\vec{p}_1^{\,\prime})^{2}
=\frac{(\vec{p}_1^{\,2}-\vec{p}_1^{\,\prime 2})^{2}}{4m}
-\vec{k}^{2}\nonumber
\end{eqnarray}
\begin{eqnarray}
\frac{1}{\sqrt{2p_{0}}}u_{(p,\lambda)}=\sqrt{\frac{m+p_{0}}{2p_{0}}}\left(\begin{array}{c}\chi(\lambda)
\\
\frac{\vec{p}.\vec{\sigma}}{m+p_{0}}\chi(\lambda)
\end{array}\right)\simeq
\left(\begin{array}{c}(1-\frac{\vec{p}^{2}}{4m^{2}})\chi(\lambda)
\\ 
\frac{1}{2m}\vec{p}.\vec{\sigma}\chi(\lambda)\end{array}\right) \label{e4}
\end{eqnarray}
Hence in the strict non-relativistic limit (i.e. $p_{0}=m$,
$1-\frac{\vec{p}^{2}}{4m^{2}}=1$) T becomes
\begin{eqnarray}
T=\frac{e^{2}}{4\pi^{2}}\chi^{\dagger}(\lambda^{'}_{1})\chi(\lambda_{1})
\frac{1}{|\vec{k}|^{2}-m_{n}^{2}}\chi^{\dagger}(\lambda^{'}_{2})
\chi(\lambda_{2})~~~~~~
|\vec{k}|<|\vec{R}|=R \label{e5}
\end{eqnarray}
\begin{eqnarray}
\gamma_{k}=\left(\begin{array}{cc}0&-\sigma_{k} \\
\sigma_{k}&0\end{array}\right)~~~~,~~~~~
\gamma_{0}=\left(\begin{array}{cc} I&0 \\
0&-I\end{array}\right)\nonumber
\end{eqnarray}
where we have introduced the cut-off R which should be taken in
the order of the momentum corresponding to the binding energy of
the nucleus. This cut-off is explicitly written in Eq.(\ref{e5}) because
$k^{2}\simeq -|\vec{k}|^{2}$ is not enough to indicate
that T in (\ref{e5}) is the non-relativistic expression since the photon
is off-shell in the propagator and one may take $k^{2}\simeq
-|\vec{k}|^{2}$ for relativistic values of
$|\vec{k}|$ as well provided that $k_{0}<<
|\vec{k}|$. In other words the strict non-relativistic
limit implies $k^{2}=-|\vec{k}|$ but $k^{2}\simeq
-|\vec{k}|^{2}$ does not necessarily imply the strict
non-relativistic limit. Therefore the explicit expression of the
conserve is not true that is
$|\vec{k}|<|\vec{R}|$ is necessary. In
non-relativistic quantum mechanics the scattering amplitude for
the elastic scattering of a particle from a potential V, in the
Born approximation may be written as \cite{Merzbacher,Sakurai,Yndurain2}
\begin{eqnarray}
T(\vec{k})=\frac{1}{(2\pi)^{2}}\int
d^{3}\vec{x}e^{-i\vec{k}\vec{x}}\chi^{\dagger}
(\lambda^{'}_{1})\chi^{\dagger}(\lambda^{'}_{2})V(\vec{x})
\chi(\lambda_{1})\chi(\lambda_{2}) \label{e6}
\end{eqnarray}
After comparing (\ref{e5}) and (\ref{e6}) one notices that
\begin{eqnarray}
f(|\vec{k}|)=\int
d^{3}\vec{x}e^{-i\vec{k}\vec{x}}V{(\vec{x})} \label{e7}
\end{eqnarray}
where
\begin{eqnarray}
f(|\vec{k}|)=\begin{array}{c}\frac{e^{2}}{|\vec{k}|^{2}-m_{n}^{2}} 
~~~~~~~~\mbox{for}~~~~ |\vec{k}|\leq R \\
0~~~~~~~~~~~~~~~\mbox{elsewhere}
\end{array}
\label{e8}
\end{eqnarray}
$V(\vec{x})$ is
obtained as the Fourier transform of
$f(|\vec{k}|)$ as
\begin{eqnarray}
V(\vec{x})&=&\frac{e^{2}}{(2\pi)^{3}}\int 
d^{3}\vec{k}\frac{e^{i\vec{k}\vec{x}}}
{|\vec{k}|^{2}-m_{n}^{2}}
=\frac{e^2}{(2\pi)^3}\int_{0}^{R}\frac{|\vec{k}|^{2}dk}{|\vec{k}|^{2}
-\frac{n^{2}}{L^{2}}}\int^{\pi}_{0}\exp\{i|\vec{k}|r\cos\theta\}\sin\theta\
d\theta\int_{0}^{2\pi}d\phi\nonumber \\
&=&\frac{e^2}{2i(2\pi)^2r}\int_{-R}^{R}\frac{kdk}
{k^{2}-\frac{n^{2}}{L^{2}}}\{\exp(ikr)-\exp(-ikr)\} \nonumber \\
&=&\frac{e^2}{i(2\pi)^2\,r}\int_{-R}^{R}
\frac{k.\exp(ikr)}{k^{2}-\frac{n^{2}}{L^{2}}}dk  \label{e9}
\end{eqnarray}
We take the wave function of two protons inside a nucleus be
\begin{eqnarray}
\Psi=\frac{\sqrt{m^{3}_{\pi}}}{\sqrt{\pi}}e^{-m_{\pi}r} \label{e10}
\end{eqnarray}
where $m_\pi$ denotes the mass of pions. Then the decay width is obtained 
as
\begin{eqnarray}
\Gamma=Im\langle\Psi|V(r)|\Psi\rangle \label{e11}
\end{eqnarray}
The evaluation of $\Gamma=\langle\Psi|V(r)|\Psi\rangle$ is done in the 
appendix and found to be
\begin{eqnarray}
\langle\Psi|V(r)|\Psi\rangle&=&i\frac{e^2\,m^3_\pi}{\pi^2}[\frac{2m\beta}{(m^{2}
-\beta^{2})^{2}}\ln(\frac{\beta+R}{\beta-R})
-\frac{(m^2+\beta^2)}{(m^{2}-\beta^{2})^2}\ln(\frac{m+R}{m-R}) 
\nonumber 
\\
&-&\frac{2mR}{(m^{2}-\beta^{2})(m^{2}-R^{2})}] 
\label{e13}
\end{eqnarray}
where $m=2im_{\pi}$, $\beta=\frac{n}{L}$. One notices that 
$\langle\Psi|V(r)|\Psi\rangle$ is real if $\beta \,>\,R$ (which is the 
most natural choice). Otherwise it means that the tachyonic photon masses 
are in the order of $MeV$. (In fact one obtains the result of \cite{Yndurain} 
when one lets $R\rightarrow\infty$.)
In other words the tachyonic photon modes can not 
lead to decay of proton through processes given in Fig.1 unless the size 
of the extra dimension is larger 
than nuclear sizes. However this does not imply that tachyonic modes can 
not induce spontaneous decay of protons once the size(s) of extra 
time-like dimension(s) are taken smaller than the nuclear sizes. There are
other contributions which may induce spontaneous decay of protons 
although the size(s) of the extra time-like dimension(s) are taken smaller 
than nuclear sizes. Such a possible contribution is induced through 
fermion self-energies as discussed in the paragraph after the next 
paragraph. An inspection of Eq.(\ref{e14}) reveals that the rate of 
spontaneous decay of a proton (or quark) is much larger than the one would 
be induced by the process given in Fig.1. Moreover fermion self-energy 
diagrams would induce an imaginary part for the pion self-energy hence 
for its mass. This, in turn, would make the pion mass in Eq.(\ref{e13}) 
complex. So there would be an imaginary contribution to Eq.(\ref{e13}) 
even in the case $R<\beta$, that is, even in the case the size of the 
extra dimension(s) are much smaller than nuclear sizes. So we will impose 
conservation of momentum in extra time-like dimensions in the paragraph 
after the next paragraph to forbid fermion self-energy diagrams with tachyonic 
photons. In that way the processes similar to Fig.1 are 
forbidden as well as the processes as in Fig.2. One may question if 
the calculation of that process (given in Eq.(\ref{e13}) ) is unnecessary 
or redundant once conservation of momentum is imposed in extra dimensions. 
In fact it is not. The result of (\ref{e13}) gives more 
flexibility in model building. For example, one may consider a process 
similar to the one given in Fig.1, where one of the incoming and outgoing 
protons are replaced by their tachyonic Kaluza-Klein counterparts. ( 
These modes may be produced in early universe in models where quarks are 
allowed to propagate in the extra time-like 
dimensions). Such processes are not forbidden by conservation of 
momentum ( in extra time-like dimensions) and their decay would be the 
same form as Eq.(\ref{e13}) provided that the wave functions for protons and 
their tachyonic counterparts have the same form as (\ref{e10}). So the 
reality of (\ref{e13}) is important in the discussion of the stability 
of protons in the presence of tachyonic modes. 

One might think that 
the scattering of high energy free protons ( e.g. in 
cosmic rays ) through processes similar to the one given in Fig.1 may 
change the bound given above. The cross section in that case can be 
directly found from (\ref{e5}) and is seen to be real. 
So the decay width of a two free nucleon system due to a process 
similar to fig.1 is zero. One may notice this fact without doing 
the 
calculation of the corresponding decay width explicitly. The decay width, 
i.e. the imaginary part of  
$\langle\Psi|V(r)|\Psi\rangle$ is due to the mixture of the 
arguments of the real 
exponent in $\Psi$ and the complex exponential in $V(r)$. If one takes 
$\Psi$ be wave function of two free protons ( which is expressed by a 
complex exponential) then in the evaluation of 
$\langle\Psi|V(r)|\Psi\rangle$ the overall complex exponentials 
cancel and 
$\langle\Psi|V(r)|\Psi\rangle$ results in a 
real number so it has no imaginary component. In other words the decay 
width of two free protons due to tachyonic photon modes is always zero.
However for confined particles one may expect a wave function of the form 
of (\ref{e10}), which results in a non-zero decay width. Hence the quarks 
inside the nucleons may give such a non-zero decay width. On the other 
hand we do not know the wave functions of quarks inside nucleons so it is 
impossible to obtain an exact lower bound on the size of extra time-like 
dimensions by 
considering the quarks inside nucleons. However one may expect this wave 
function not be drastically different from (\ref{e10}). In that case one 
would expect the lower bound on the size of extra dimensions be in the 
order of ( cut-off momentum)$^{-1}$, that is, 
${\cal O}(\frac{1}{1GeV})$. 
In the same way one may put a still smaller lower limit if quarks are made 
of composites of some other particles (preons). If this generalization is 
reliable then one may relate the lower limit on the size of extra 
time-like 
dimension(s) and the binding energy. In this case one may speculate that, 
if an extra time-like 
dimension of the size much larger than the (Planck mass)$^{-1}$ is 
discovered 
then it excludes possibility of stable bound states with energies much 
higher than the inverse of the size of the extra time-like dimension.  

Next we consider the problem of the spontaneous decay of a particle (e.g. 
electron) into a tachyon and the original particle, and the problem of 
imaginary mass contribution to the stable fermions (e.g. electron 
or proton) through self energy diagrams involving a tachyon. The decay of 
a particle ( say an electron) into another electron and a 
negative energy tachyonic photon is kinematically allowed.
It is difficult to identify these negative energy tachyons 
with anti-tachyons because negative energy tachyons may be made positive 
energy by a simple Lorentz boost \cite{Feinberg}. So the result of such 
decays can be 
catastrophic because the kinematics allows large negative values for 
the energy of such a tachyon and such a large negative value 
energy destabilizes the whole vacuum.  
However once we identify the tachyon with the Kaluza-Klein mode of the 
photon in the extra time dimension this decay becomes impossible since (at 
least in the transient time till the formation of the 
standing waves) there will be a non-zero net 
momentum flow in the extra time direction due to the tachyon and there is 
no other momentum to balance it. 
\begin{figure}[*h]
\vspace*{.50in}
\hspace*{-0.7in}
\begin{minipage}{4.75in}
\begin{center}
\epsfig{file=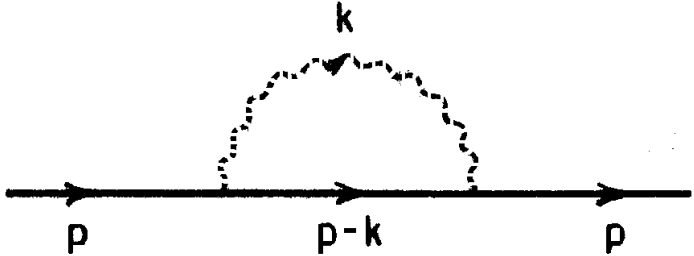,height=1.35in}
\end{center}
\end{minipage}
%\vskip .2in
\caption{\label{fig2}
{\it
The Feynman diagram for the contribution of a photonic tachyon to fermion 
self-energy. The wavy 
line denotes the tachyonic
Kaluza-Klein modes of photon and the solid line denotes the fermion}} 
\end{figure}
%\clearpage
The problem of the imaginary 
contribution to the masses of stable fermions through self-energy diagrams 
involving tachyons can be avoided in the same way i.e. by 
imposing the conservation of momentum corresponding to the extra time-like 
dimension. Without taking this conservation into account, the contribution 
of the self-energy diagram given in Fig.2 to the fermion mass  
(in the Pauli-Villars regularization scheme) is of the form
\begin{equation}
\delta m \propto 
\frac{e^2\,m}{4\pi^2}\ln{\frac{\mu^2-\Lambda^2}{\mu^2}}
~~~~,~~~~\mu^2>0 \label{e14}
\end{equation}
where $m$, $e$, $\mu$, $\Lambda$ stand for the fermion mass, the electric 
charge of the fermion, the mass of the tachyonic photon, 
the Pauli-Villars regularization 
cut-off scale; respectively and we have modified the 
propagator of the tachyonic photon mode ( in the Pauli-Villars 
regularization scheme) by 
\begin{equation}
\frac{1}{k^2-\mu^2}\rightarrow 
(\frac{1}{k^2-\mu^2})\frac{\Lambda^2-\mu^2}{k^2-\mu^2+\Lambda^2} 
\label{e15}
\end{equation} 
By definition $\Lambda\,>\,\mu$ so Eq.(\ref{e14}) 
results 
in an imaginary contribution of the form 
\begin{equation}
i\frac{e^2\,m}{4\pi} \label{e16}
\end{equation}
which is independent of $\mu$ and $\Lambda$ and essentially equal to the 
width of the spontaneous decay of the fermion through release of a 
tachyonic photon. This result is extremely problematic because it predicts 
a decay rate for the fermion comparable to the decay width of 
hadronic resonances and moreover the result in Eq.(\ref{e16}) may be 
multiplied by a large number because the number of 
Kaluza-Klein modes is about $\frac{\Lambda}{\mu_0}$ where $\mu_0$ is the 
mass of the first Kaluza-Klein mode and $\Lambda$ is at most at the 
order of Planck mass. However if we require conservation of the momentum in the 
extra 
time direction (at least in the transient time till the formation of 
standing waves) then usual fermions (i.e. Kaluza-Klein zero modes of 
fermions) 
can only 
radiate usual photons (i.e. Kaluza-Klein zero modes of photons) and the 
contribution to the 
fermion self-energies
given by Fig.2 is absent and hence the problem is removed. In other words 
the contribution of a tachyonic photon to the electron mass ( as given in 
Fig.2) results in extremely problematic results if the tachyonic mode is 
not due to an extra time dimension. On the other hand the diagram in Fig.2 
is forbidden ( hence the problem is removed) if one considers the tachyon 
be due to an extra time dimension and require the conservation of momentum 
corresponding to this dimension.

In this study we have re-examined some phenomenological 
difficulties due to tachyonic photon modes in the study of extra time-like 
dimension(s). We have shown that the lower bound on the size of extra time 
dimension(s) due to the lower bound on the lifetime of proton may be 
relaxed and the presence of tachyons 
related to the extra time dimension(s) is not as problematic as the 
tachyons in the usual 4-dimensional picture. 
Although we believe that we have made some 
progress in the phenomenological viability of extra time-like 
dimensions there are still some points to be studied further. We hope that 
this study will facilitate 
more freedom in model building in future studies.

\begin{acknowledgments} We would like to thank Dr. Gregory Gabadadze for 
his valuable comments and suggestions.
\end{acknowledgments}

\appendix
\section{}
In this appendix we give the details of the evaluation of the integral 
given in (\ref{e13}). 
\begin{eqnarray}
&&\int_{-R}^{R}\frac{k
dk}{k^{2}-\frac{n^{2}}{L^{2}}}\int_{0}^{\infty}re^{(ik-2m_{\pi})r}dr
\int_{0}^{\pi}\sin\theta
d\theta\int_{0}^{2\pi}d\phi
=4\pi\int_{-R}^{R}\frac{k
dk}{k^{2}-\frac{n^{2}}{L^{2}}}\int_{0}^{\infty}
re^{(ik-2m_{\pi})r} dr\nonumber \\
&&=-4\pi\,\int_{-R}^{R}\frac{k
dk}{(k^{2}-\frac{n^{2}}{L^{2}})(k+2\,i\,m_{\pi})^{2}}
\label{a1}
\end{eqnarray}
The denominator of the integral may be written as
\begin{eqnarray}
\frac{1}{(k+\beta)(k-\beta)(k+m-\epsilon)(k+m+\epsilon)}
=\frac{1}{(x-x_{1})(x-x_{2})(x-x_{3})(x-x_{4})}\label{a2}
\end{eqnarray}
where
\begin{eqnarray}
&&m=2im_{\pi}~,~~~ \beta=\frac{n}{L} \nonumber \\
&&x=k,~~~x_{1}=-\beta,~~~x_{2}=\beta,~~~x_{3}=-m+\epsilon,
~~~x_{4}=-m-\epsilon \label{a3}
\end{eqnarray}
We use the identity
\begin{eqnarray}
\frac{1}{(x-x_{1})(x-x_{2})}=\frac{1}{x_{1}-x_{2}}[\frac{1}{x-x_{1}}-\frac{1}{x-x_{2}}]
\end{eqnarray}
to write (\ref{a2}) as
\begin{eqnarray}
&&\frac{1}{(x-x_{1})(x-x_{2})(x-x_{3})(x-x_{4})} \nonumber \\
&=&\frac{1}{x_{1}-x_{2}}\{\frac{1}{(x_{1}-x_{3})(x_{1}-x_{4})}.
\frac{1}{x-x_{1}}-\frac{1}{(x_{2}-x_{3})(x_{2}-x_{4})}.\frac{1}{x-x_{2}}\}
\nonumber \\
&&+\frac{1}{x_{3}-x_{4}}\{\frac{1}{(x_{1}-x_{3})(x_{2}-x_{3})}
\frac{1}{x-x_{3}}-\frac{1}{(x_{1}-x_{4})(x_{2}-x_{4})}\frac{1}{x-x_{4}}\}
\label{a4}
\end{eqnarray}
The second term in (\ref{a4})is 
\begin{eqnarray}
&&\frac{1}{x_{3}-x_{4}}\{\frac{1}{(x_{1}-x_{3})(x_{2}-x_{3})}
\frac{1}{x-x_{3}}-\frac{1}{(x_{1}-x_{4})(x_{2}-x_{4})}\frac{1}{x-x_{4}}\} 
\nonumber \\
&=&
\frac{1}{2\epsilon}\{a\frac{1}{x-x_{3}}-b\frac{1}{x-x_{4}}\}
=\frac{1}{2\epsilon}\{\frac{(a-b)x+bx_{3}-ax_{4}}{(x-x_{3})(x-x_{4})}\}
\label{a5}
\end{eqnarray}
where
\begin{eqnarray}
&&a=\frac{1}{(x_{1}-x_{3})(x_{2}-x_{3})}~,
~~~~~b=\frac{1}{(x_{1}-x_{4})(x_{2}-x_{4})} \label{a61} \\
&&
\frac{(a-b)x}{x_{3}-x_{4}}=\frac{2mx}{[(m-\epsilon)^{2}-\beta^{2}]
[(m+\epsilon)^{2}-\beta^{2]}} \label{a62} \\
&&
\frac{bx_{3}-ax_{4}}{x_{3}-x_{4}}=-\frac{\beta^{2}-3m^{2}-\epsilon^{2}}
{[(m-\epsilon)^{2}-\beta^{2}][(m+\epsilon)^{2}-\beta^{2}]} 
\label{a63}
\end{eqnarray}
then (\ref{a5}) becomes
\begin{eqnarray}
&&\frac{1}{x_{3}-x_{4}}\{\frac{1}{(x_{1}-x_{3})(x_{2}-x_{3})}
\frac{1}{x-x_{3}}
-\frac{1}{(x_{1}-x_{4})(x_{2}-x_{4})}\frac{1}{x-x_{4}}\}\nonumber \\
&&=\{\frac{2mx}{[(m-\epsilon)^{2}-\beta^{2}]
[(m+\epsilon)^{2}-\beta^{2}]}
\nonumber \\
&&-\frac{\beta^{2}-3m^{2}-\epsilon^{2}}
{[(m-\epsilon)^{2}-\beta^{2}][(m+\epsilon)^{2}-\beta^{2}]}\}
\frac{1}{(x-x_{3})(x-x_{4})} 
\label{a7}
\end{eqnarray}
After combining (\ref{a5}), (\ref{a7}) and using the explicit values 
of $x_1$, $x_2$, $x_3$, $x_4$; and letting $\epsilon\rightarrow\,0$ one 
obtains
\begin{eqnarray}
\frac{k}{(k^{2}-\beta^{2})(k+m)^{2}}&=&-\frac{1}{2\beta(m-\beta)^{2}}
\frac{k}{k+\beta}+\frac{1}{2\beta(m+\beta)^{2}}\frac{k}{k-\beta}\nonumber 
\\
&+&\frac{2m}{(m^{2}-\beta^{2})^{2}}\frac{k^{2}}{(k+m)^{2}}-
\frac{\beta^{2}-3m^{2}}{(m^{2}-\beta^{2})^{2}}\frac{k}{(k+m)^{2}}\label{a8}
\end{eqnarray}
The evaluation of the integral (\ref{a1}) by the use of  
(\ref{a8}) gives
\begin{eqnarray}
\langle\Psi|V(r)|\Psi\rangle=&&i\frac{e^2\,m_\pi^3}{\pi^2}[\frac{2m\beta}{(m^{2}
-\beta^{2})^{2}}\ln(\frac{\beta+R}{\beta-R})
-\frac{(m^2+\beta^2)}{(m^{2}-\beta^{2})^2}\ln(\frac{m+R}{m-R}) \nonumber 
\\
&-&\frac{2m\,R}{(m^{2}-\beta^{2})(m^{2}-R^{2})}]
\label{a9}
\end{eqnarray}

\bibliographystyle{plain}

\end{document}